# Facilitating bootstrapped and rarefaction-based microbiome diversity analysis with q2-boots

Isaiah Raspet[1,2], Elizabeth Gehret[1], Chloe Herman[1,2], Jeff Meilander[1], Andrew Manley[1], Anthony Simard[1], Evan Bolyen[1], J. Gregory Caporaso[1,2,3]

1. Center for Applied Microbiome Science, Pathogen and Microbiome Institute, Northern Arizona University, Flagstaff, AZ, USA.
2. School of Informatics, Computing and Cyber Systems, Northern Arizona University, Flagstaff, AZ, USA
3. Corresponding author: greg.caporaso@nau.edu

# Abstract

**Background**: We present q2-boots, a QIIME 2 plugin that facilitates bootstrapped and rarefaction-based microbiome diversity analysis. This plugin provides eight new actions that allow users to apply any of thirty different alpha diversity metrics and twenty-two beta diversity metrics to bootstrapped or rarefied feature tables, using a single QIIME 2 Pipeline command, or more granular QIIME 2 Action commands.

**Results**: Given a feature table, an even sampling depth, and the number of iterations to perform (`n`), the command `qiime boots core-metrics` will resample the feature table `n` times and compute alpha and beta diversity metrics on each resampled table. The results will be integrated in summary data artifacts that are identical in structure and type to results that would be generated by applying diversity metrics to a single table. This enables all of the same downstream analytic tools to be applied to these tables, and ensures that all collected data is considered when computing microbiome diversity metrics.

**Conclusions**: A challenge of this work was deciding how to integrate distance matrices that were computed on `n` resampled feature tables, as a simple average of pairwise distances (median or mean) does not account for the structure of distance matrices. q2-boots provides three options, and we show here that the results of these approaches are highly correlated. q2-boots is free and open source. Source code, installation instructions, and a tutorial can be found at https://github.com/caporaso-lab/q2-boots.

# Background

Two recent papers [1, 2] have brought discussion about rarefaction back into focus in microbiome diversity analysis. Inspired by this work we developed q2-boots, a new QIIME 2 [3] plugin that facilitates bootstrapped and rarefaction-based diversity analysis for microbiome researchers and data scientists. While rarefaction analysis has always been possible in QIIME 2, the diversity analysis workflow that most users consider the default (i.e., running `qiime diversity core-metrics`, or `qiime diversity core-metrics-phylogenetic`) only includes a single step of rarefying the input feature table to a user-specified sequencing depth. Other workflows, such as `qiime diversity alpha-rarefaction` and `qiime diversity beta-rarefaction`, enable rarefaction analysis (i.e., multiple iterations of rarefying the input feature table and comparison of the diversity metrics computed on each of the rarefied tables), but because these result in terminal QIIME 2 Visualizations, rather than QIIME 2 Artifacts that can be used in downstream steps such as statistical analysis, performing analysis on rarefaction data in QIIME 2 is currently not common in practice.

# Implementation

q2-boots was designed to serve as a stand-in replacement for the widely used q2-diversity plugin. It provides eight new Actions (i.e., `qiime2.Action` objects) that broadly fall into four categories: *resampling,* the *alpha diversity suite* of Actions, the *beta diversity suite* of Actions, and the *core metrics* Pipeline (Figure 1).

The computationally expensive steps of the q2-boots plugin (i.e., resampling tables, and diversity calculations on the many resulting tables) can all be run in parallel if requested by the user, supported by QIIME 2's use of parsl [4] for its formal parallel computing support. This enables full use of individual multiprocessor computers (such as laptops or cloud-based virtual machines such as those available from Amazon Web Services or Digital Ocean) or multi-node high performance computing machinery (e.g., university cluster computers). Additionally, by nature of being a QIIME 2 plugin, failed Pipeline runs can be resumed avoiding the need to recompute results. This is useful, for example, if an out-of-memory error requires a re-run of analysis with a larger memory allocation after a large portion of work has already been completed.

As we discuss implementation details in the following subsections we use some terms that have specific meaning in QIIME 2, such as Action, Pipeline, and Artifact Class. The glossary of *Developing with QIIME 2* [5] is the canonical reference for definitions of these terms.

## Resampling

A single Action, `resample`, supports resampling an input feature table (QIIME 2 artifact class: `FeatureTable[Frequency]`) to a user-specified per-sample sampling depth `n` times. Sampling is performed with replacement (i.e., bootstrapped) or without replacement (i.e., rarefied [2]). Samples with total frequencies less than the specified sampling depth are dropped from the resulting feature tables. The output of this Action is `n` feature tables that can be used in downstream QIIME 2 Actions, or which can be exported to biom-formatted files [6] for analysis with other tools.

In a comparison of bootstrapped and rarefaction-based diversity analysis on a collection of 1,018 samples spanning human excrement, compost, and soil samples from a currently unpublished longitudinal microbiome analysis of a human excrement composting (HEC) experiment, we find that the results are effectively identical when comparing averaged alpha and beta diversity metrics computed from the resulting tables (Supplementary Table 1). In two iterations each of bootstrapped and rarefaction-based diversity analysis, 100 resampled tables were created. Four alpha and four beta diversity metrics were computed on each resampled table, and the resulting alpha diversity vectors and beta diversity distance matrices were averaged. On a per-diversity metric basis, correlation between the averaged alpha diversity vectors or beta diversity distance matrices resulting from bootstrapped or rarefaction-based resampled feature tables, measured with Spearman rank correlation and Mantel test using Spearman rank correlation respectively, all achieved correlation coefficient (rho) values greater than 0.99 and p-values less than 0.001. These results, and all code used to perform the analysis, are presented in Supplementary Data.

Rarefaction-based diversity analysis is more common than bootstrapped diversity analysis in microbiome research, but has its roots in macro-ecology, where the population sizes being sampled from are small relative to those sampled in microbial ecology. The availability of both of these approaches through a common interface will facilitate assessment of whether one is more appropriate in microbiome diversity analysis generally, or under specific circumstances.

## Alpha diversity suite

The *alpha diversity suite* introduces three Actions: `alpha`, `alpha-collection`, and `alpha-average`. `alpha` is an analog of q2-diversity's `alpha` and `alpha-phylogenetic` actions. Those Actions were introduced in q2-diversity before optional input Artifacts were supported in QIIME 2, so two Actions were required to support phylogenetic diversity metrics (which take a feature table and a phylogenetic tree as input) and non-phylogenetic diversity metrics (which only take a feature table as input). The implementation of `alpha` in q2-boots improves upon this by making the phylogenetic tree an optional input, and therefore supports both phylogenetic and non-phylogenetic diversity metrics. This same change relative to q2-diversity was made for the *beta diversity suite* of Actions and *core metrics*. (Because the QIIME 2 Amplicon Distribution strives to maintain backward compatibility, q2-diversity has not been updated to use optional inputs for these Actions.)

The implementation of `alpha` in q2-boots integrates resampling of the input feature table `n` times. In addition to providing the typical inputs and parameters for an alpha diversity computation (a feature table, an optional phylogenetic tree, and the name of a supported diversity metric to compute), the user provides an even sampling depth, whether to sample with or without replacement, and the number of resampled feature tables to compute when calling `qiime boots alpha`. These parameters are passed through to the `resample` Action described above. The input feature table is resampled `n` times to create `n` feature tables, the user-specified alpha diversity metric is computed on all samples in each feature table, the per sample diversity metrics computed on each table are averaged (the averaging method is median, by default), and the Action outputs a single vector of the averaged per-sample alpha diversity metric values. The resulting vector can be used in any downstream actions compatible with this type (QIIME 2 artifact class: `SampleData[AlphaDiversity]`) in the QIIME 2 ecosystem, or exported as tab-separated text for use elsewhere. This facilitates the use of bootstrapping or rarefaction in alpha diversity analysis.

The `alpha-collection` Action works similarly to the `alpha` Action, but rather than integrating the averaging step, it outputs the full collection of alpha diversity vectors that were computed on each of the resampled feature tables. It thus allows users to interact directly with these diversity vectors to support estimates of variance across resampled tables, or other computations that require the full data rather than simple averages.

The final tool in the alpha suite, `alpha-average`, takes a collection of alpha diversity vectors, such as those generated by `alpha-collection`, computes their average (the default averaging method is median) and outputs a single vector containing per-sample averages.

`alpha` is implemented as a Pipeline (i.e., a `qiime2.Pipeline` object) that integrates the `resample`, `alpha-collection`, and `alpha-average` Actions into a single command. The availability of the Pipeline facilitates the use of this workflow by all QIIME 2 users. The availability of the component Actions (`alpha-collection` and `alpha-average`) provides flexibility for more advanced users who may need access to the intermediary data.

## Beta diversity suite

The beta diversity suite of Actions in q2-boots mirrors the Actions in the alpha diversity suite. `beta` is the analog to q2-diversity's `beta` and `beta-phylogenetic` Actions, integrating bootstrapping or rarefaction and outputting a distance matrix (QIIME 2 artifact class: `DistanceMatrix`). Like `alpha`, `beta` is a Pipeline composed of calls to `resample`, `beta-collection`, and `beta-average`. `beta-collection` outputs the collection of distance matrices computed on each resampled table, and `beta-average` averages those to create a single distance matrix that can be used in any downstream analysis that operates on distance matrices in the QIIME 2 ecosystem, or exported as tab-separated text for analysis with other tools.

Averaging distance matrices is more complex than averaging alpha diversity vectors because distance matrices are observations within a metric space with specific properties: 1. Hollowness (i.e., the diagonal of the distance matrix must be zero); 2. Non-negative (i.e., all values must be greater than or equal to zero); 3. Symmetry (i.e., the values must be symmetric across the diagonal, such that the distance between samples A and B is always equal to the distance between samples B and A); and 4. Triangle Inequality (e.g., the distance between sample A and C is always greater than or equal to the distance between samples A and B plus the distance between samples B and C).

`beta-average`, and all other Actions that use it in q2-boots, provides three options for averaging distance matrices: `medoid`, which retains all of these properties (assuming the distance metric used produces distance matrices with these properties), and `non-metric-median` and `non-metric-mean`, which retain properties 1-3, but may not retain property 4.

`medoid` takes a set of distance matrices, identifies the distance matrix with the smallest sum of dissimilarities (by Euclidean distance) to all of the other distance matrices in the set, and returns that distance matrix as a representative of the set. Thus if all distance matrices in the set are metric, the selected representative will also be metric. `non-metric-median` and `non-metric-mean` are naive approaches that take a set of distance matrices and compute the per-cell median or mean, respectively, across the set of distance matrices. The result is a new distance matrix composed of the averaged pairwise distances.

q2-boots uses the medoid implementation from hdmedians (https://github.com/daleroberts/hdmedians, accessed 9 May 2024). In practice, we have found that the memory requirements of this method don't scale well to large numbers of `n` (e.g., `n` > 100). To assess whether non-metric-median or non-metric-mean is a suitable surrogate for medoid in lieu of a more memory efficient medoid implementation, we computed distance matrices for the HEC experiment described above using four beta diversity metrics. We find that on a per-distance-metric basis, the distance matrices resulting from applying `qiime boots beta-average` with the medoid, non-metric-mean, and non-metric-median averaging methods across 100 bootstrap and rarefaction iterations are effectively identical (Supplementary Table 1). All Mantel tests achieved correlation coefficients greater than 0.99 and p-values less than 0.001. We additionally compared multiple iterations of bootstrapping and rarefaction followed by averaging of the resulting distance matrices with `medoid`, `non-metric-median`, and `non-metric-mean`, to assess whether different random seeds during resampling would impact the similarity of the resulting distance matrices. Using the same four distance metrics, we again found that all Mantel rho scores were greater than 0.99 and all p-values were less than 0.001. These results, and all code used to perform the analysis, are presented in Supplementary Data.

## Core metrics

q2-boots provides an omnibus Action, `core-metrics`, that captures the behavior of q2-diversity's `core-metrics` and `core-metrics-phylogenetic` Actions. Like the `alpha` and `beta` Actions, `core-metrics` integrates bootstrapping and rarefaction. The output it creates mirrors that of the corresponding Actions in q2-diversity: alpha diversity vectors that are averaged across the collection of alpha diversity vectors computed for each rarefied feature table; beta diversity distance matrices that are averaged across the collection of distance matrices computed for each rarefied feature table; principal coordinates analysis (PCoA) matrices computed on each average distance matrix; and Emperor-based interactive ordination plots [7] for each PCoA matrix. This Action enables QIIME 2 users to run bootstrapped or rarefaction-based diversity analysis with no additional work relative to running diversity analysis through the widely used `core-metrics` and `core-metrics-phylogenetic` Actions in q2-diversity.

## Testing, distribution, and maintenance

q2-boots integrates unit tests that cover the breadth of its functionality. These tests are automatically run on every commit to the main branch of the code repository, and on all pull requests.

q2-boots is available on GitHub at https://github.com/caporaso-lab/q2-boots. It is distributed as a QIIME 2 community plugin, meaning that at this time it is not included in the canonical QIIME 2 distributions, but rather is built and distributed as recommended in *Developing with QIIME 2*. As of this writing, q2-boots can be installed in the latest development version (2024.10) and release version (2024.5) of the amplicon and metagenome distributions of QIIME 2 by following instructions linked from the project's README.

q2-boots was developed for a Master's degree in Computer Science project, and will be maintained by the Caporaso Lab at Northern Arizona University. Technical support is available on the QIIME 2 Forum (https://forum.qiime2.org).

## Conclusions

q2-boots provides eight actions that facilitate bootstrapped and rarefaction-based microbiome diversity analysis workflows. The topic of rarefaction in microbiome analysis remains controversial, though as pointed out in [1], these analyses do enable consideration of all data, including low abundance features such as rare taxa. Whether using these Actions directly, as advocated in [2], or as a basis for comparison against more sophisticated normalization techniques, q2-boots makes it straightforward for researchers to integrate bootstrapped and/or rarefaction-based microbiome diversity analysis in their workflows.

# Availability and requirements

**Project name**: q2-boots
**Project home page**: https://github.com/caporaso-lab/q2-boots
**Operating systems**: Linux, macOS, Windows via Windows Subsystem for Linux
**Programming language**: Python 3
**Other requirements**: QIIME 2 2024.5 or later
**License**: BSD 3-Clause
**Any restrictions to use by non-academics**: None; q2-boots is open source and free for all use.

# Declarations

## Availability of data and material

All data analyzed in this study, including Supplementary Table 1 and the code used to generate it, are available in Supplementary Data at https://doi.org/10.5281/zenodo.13287126. Human fecal samples were collected under Northern Arizona University IRB protocol 1773199-3, *Bridging the Gap between Gut and Soil Microbiomes*.

## Competing interests

Not applicable.

## Funding

This work was funded in part by the National Cancer Institute grant 1U24CA248454-01 to JGC.

## Authors' contributions

IR and JGC were the primary developers of q2-boots. EG, CH, AS, EB, and JGC advised on the development of q2-boots. AS and EB performed code review prior to the initial release of q2-boots. JM and JGC generated the data used in Supplementary Table 1. AO and JGC performed testing of q2-boots on real-world data. EB and JGC conceived of the project. IR and JGC drafted the first version of the manuscript. All authors reviewed and provided feedback on the manuscript.

# Figure captions

**Figure 1: Workflows available in q2-boots.** q2-boots provides workflows for (a) resampling a feature table with or without replacement; (b) performing alpha diversity analysis, integrating resampling; and (c) performing beta diversity analysis, integrating resampling. An omnibus workflow, core-metrics, integrates the resampling, alpha diversity, and beta diversity workflows such that alpha and beta diversity metrics are computed on the same collection of resampled feature tables.

# Supplementary Files

Supplementary Data and Supplementary Table 1 are available at https://doi.org/10.5281/zenodo.13287126 [8]. They are not included in this pre-print due to size.

# List of works cited

# Figure 1a. Resampling workflow.

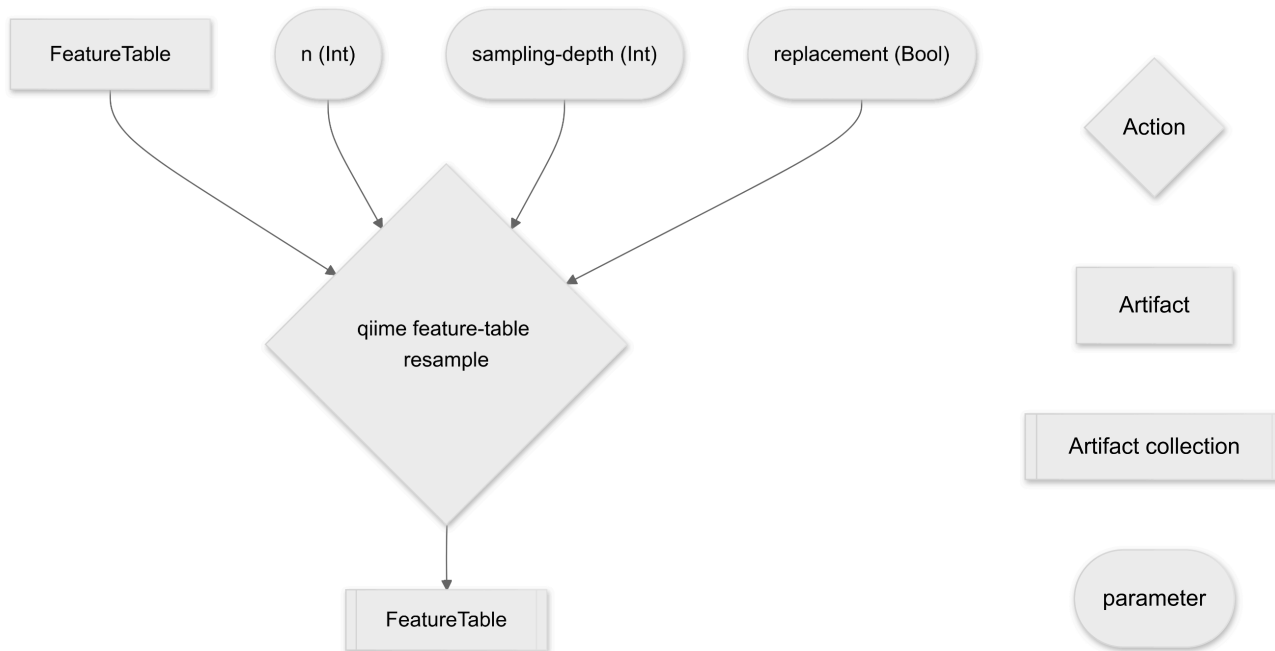

# Figure 1b. Alpha diversity workflow.

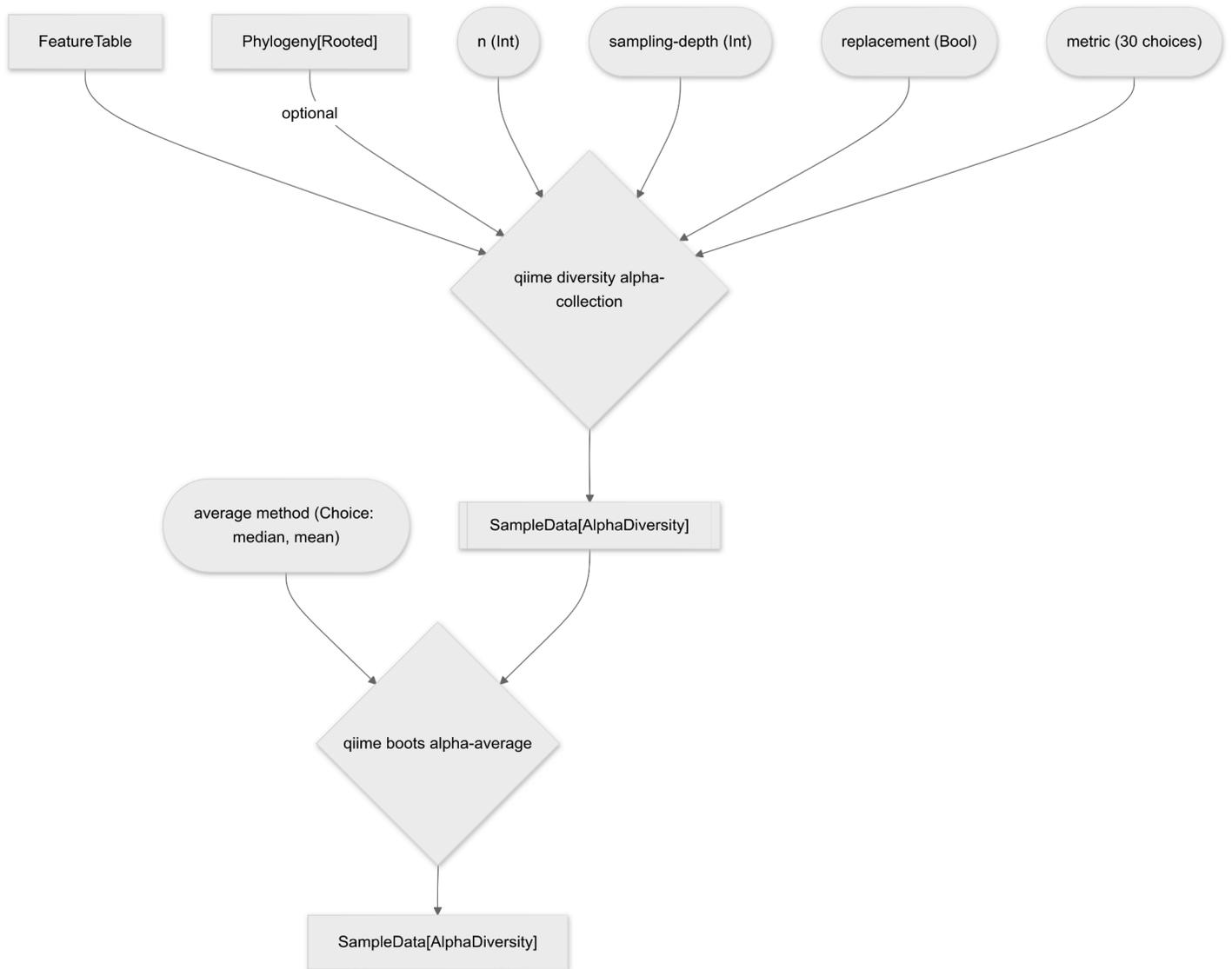

Figure 1c. Beta diversity workflow.

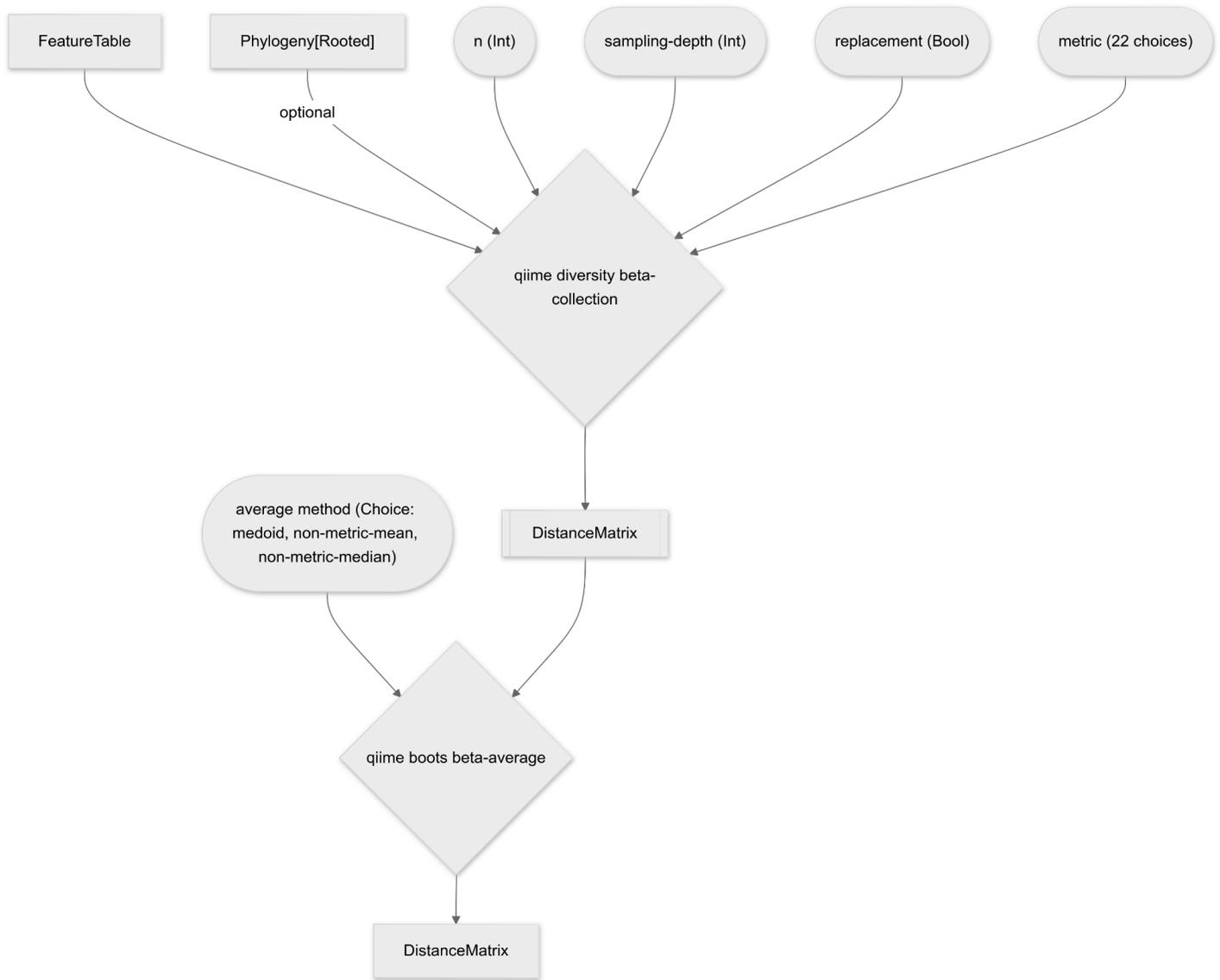